\documentclass[conference]{IEEEtran}
\IEEEoverridecommandlockouts
\usepackage{cite}
\usepackage{amsmath,amssymb,amsfonts}
\usepackage{algorithmic}
\usepackage{graphicx}
\usepackage{textcomp}
\usepackage{xcolor}
\def\BibTeX{{\rm B\kern-.05em{\sc i\kern-.025em b}\kern-.08em
    T\kern-.1667em\lower.7ex\hbox{E}\kern-.125emX}}
\begin{document}

\title{A Proof of Concept SRAM-based Physically Unclonable Function (PUF) Key Generation Mechanism for IoT Devices}
% {\footnotesize \textsuperscript{*}Note: Sub-titles are not captured in Xplore and
% should not be used}

% \author{\IEEEauthorblockN{Ashwija Reddy Korenda}
% \IEEEauthorblockA{\textit{School Of Informatics, Computing and Cyber Sytems} \\
% \textit{Northern Arizona University}\\
% Flagstaff, Arizona \\
% ashwijakorenda@nau.edu}
% \and
% \IEEEauthorblockN{Fatemeh Afghah}
% \IEEEauthorblockA{\textit{School Of Informatics, Computing and Cyber Sytems} \\
% \textit{Northern Arizona University}\\
% Flagstaff, Arizona \\
% fatemeh.afghah@nau.edu}
% \and
% \IEEEauthorblockN{Bertrand Cambou}
% \IEEEauthorblockA{\textit{School Of Informatics, Computing and Cyber Sytems} \\
% \textit{Northern Arizona University}\\
% Flagstaff, Arizona \\
% bertrand.cambou@nau.edu}
%  \and
%  \IEEEauthorblockN{Christopher Philabaum}
%  \IEEEauthorblockA{\textit{School Of Informatics, Computing and Cyber Sytems} \\
% \textit{Northern Arizona University}\\
% Flagstaff, Arizona \\
% cp723@nau.edu}
% \thanks{This material is based upon the work supported by the National Science Foundation under Grant No. 1827753. }

\author{
	\IEEEauthorblockN{Ashwija Reddy Korenda, Fatemeh Afghah, Bertrand Cambou, Christopher Philabaum}
% 	\thanks{Distribution A: Approved for Public Release, distribution unlimited.  Case Number 88ABW-2019-???? on ?? Apr. 2019. The work of F. Afghah, J. Ashdown and K. Turck was supported by AFRL. }}
	\thanks{This material is based upon the work supported by the National Science Foundation under Grant No. 1827753.\newline\newline 
	978-1-7281-2294-6/19\$31.00 \text{\textcopyright}2019 IEEE}
	\IEEEauthorblockA{School of Informatics, Computing and Cyber Systems, Northern Arizona University, Flagstaff, AZ, USA
	\\
	{\fontfamily{qcr}\selectfont
\{ashwijakorenda, fatemeh.afghah, bertrand.cambou, cp723\}@nau.edu
}}

 }
 
\maketitle
\begin{abstract}
This paper provides a proof of concept for using SRAM based Physically Unclonable Functions (PUFs) to generate private keys for IoT devices. PUFs are utilized, as there is inadequate protection for secret keys stored in the memory of the IoT devices. We utilize a custom-made Arduino mega shield to extract the fingerprint from SRAM chip on demand. We utilize the concepts of ternary states to exclude the cells which are easily prone to flip, allowing us to extract stable bits from the fingerprint of the SRAM. Using the custom-made software for our SRAM device, we can control the error rate of the PUF to achieve an adjustable memory-based PUF for key generation. We utilize several fuzzy extractor techniques based on using different error correction coding methods to generate secret keys from the SRAM PUF, and study the trade-off between the  false authentication rate and false rejection rate of the PUF. 
\end{abstract}


\section{Introduction}
Internet of things is one of the booming technologies in the current era. It allows the connection of various sensors, devices and everyday objects to interact with each other, transfer and retrieve data, intelligently respond and trigger actions accordingly. Today, many devices which paved the way to the advancement of smart homes to smart cars are connected through IoT networks.
Medical devices like pacemakers and neurosimulators which  allow monitoring of patients from afar, are connected through IoT \cite{patton2014uninvited}. %Even high tech pacemakers and neurosimulators are connected through IoT to exchange information with the hospital staff. This will allow the hospital to be in constant contact with the patients pacemakers thus allowing them to monitor the patient from afar. 
%\textcolor{magenta}{need references, but generally I think it's a long opening statement and not directly related to the objective of the paper}\textcolor{blue}{Added reference, made the statement more concise.}

Though IoT allows massive advancements in various fields, it comes with unique challenges considering its heterogeneous nature and the large number of devices. By the year 2020, it is expected that about 20 billion devices will be connected to IoT network. Sensitive information is exchanged between different IoT ``things", which is responsible for the proper functioning of smart car, pacemakers, neurosimulators or smart home. Bio-medical devices such as pacemakers and neurosimulators usually send small electric pulses to the heart and brain, respectively. If adversaries manage to send unwanted electric pulses to the brain or heart, it may lead to serious threats to patient safety. For instance, if such sensitive information is hacked or tampered, it may cause serious injuries or death of a person \cite{marin2018securing}. An article published by CNBC in 2016 discusses the possibility of cyber-criminals targeting medical devices \cite{cnbc2016}. Therefore, it is very important to ensure safe and secure encryption of messages. Success of IoT depends on the security it can provide over innovation.

% IoT devices are usually low power devices, therefore traditional security mechanisms which require constant power supply or high computational power are not suitable for IoT networks. 

Security challenges in IoT encompass different aspects of identification, authentication, encryption, confidentiality, jamming, cloning, hijacking and privacy as explained in Table \ref{tab:Definitions}. IoT networks are prone to various types of attacks such as spoofing, data manipulation, replay routing, Denial of Service (DoS), node capture and Sybil attack which are briefly explained in table \ref{tab:Attacks}
\begin{table*}[t]
\caption{Definitions of Different Security Threats in IoT Networks}
\begin{center}
\begin{tabular}{|c|c|}
\hline
\textbf{Name} & \textbf{Definition} \\ 
\hline
Identification & Unique information describing the subject, which can be utilized to identify it\\
\hline
Authentication & Secret information confirming the identity of the subject \\
\hline
Encryption & Conversion of plain text to protected message, called cipher \\
\hline
Confidentiality & State of keeping a secret \\
\hline
Jamming & Denial of Service attack in a wireless medium \\ 
\hline
Cloning & Replicating the process\\
\hline
Hijacking & Attacker gaining control of the system\\
\hline
Privacy & Ability to protect sensitive information\\
\hline

\end{tabular}
\label{tab:Definitions}
\end{center}
\end{table*}

\begin{table*}[t]%[!htbp]
\centering
\caption{Attacks in IoT networks}
\label{tab:Attacks}

\begin{tabular}{|c|c|}
\hline
\textbf{Attack}  & \textbf{Impact}  \\
\hline
Spoofing & Malicious user impersonating another device to launch attacks on network hosts\\
\hline
Replay routing information & Fraudulent repeating or delaying of valid data transmission \\
\hline
Data manipulation & Altering sensitive information \\
\hline
Denial of Service (DoS) & Reduction in networks' capacity, disable the network \\
\hline
Node capture attack & Stealing sensitive information from a captured node to compromise the entire network \\
\hline
Sybil attack & Node claiming multiple identities \\
\hline
\end{tabular}

\end{table*}

Several characteristics of IoT networks including the low power of IoT nodes, the low computational capability combined with the heterogeneity and large scale nature of IoT networks limit the application of standard security mechanisms to the IoT networks. In this paper, we focus on the problem of identification, authentication and encryption of the messages sent between two IoT devices to protect the confidentiality of the sensitive information transmitted.

Encryption has been widely used by several mechanisms in order to send the messages without the risk of being understood by the hackers and to eliminate the risk of data manipulation. Hence, cryptographic methods play a crucial role in the security of IoT systems.
The majority of common cryptographic mechanisms such as public key infrastructure (PKI), advanced encryption standard (AES), and elliptic curve cryptography (ECC) rely on private cryptographic keys \cite{dodis2004fuzzy}. These secret cryptographic keys are expected to be robust, reliable and reproducible and there are usually stored in the non-volatile memory (NVM) of the devices. However, the NVMs are highly susceptible to physical attacks due to their robust electrical nature. Several hardware-based security solutions have been developed to enhance the security of private-key based cryptographic methods \cite{kinney2006trusted,barker2011transitions}. 

Physically unclonable functions (PUFs) are a hardware-based security primitive introduced in 2002 \cite{gassend2002silicon}. The PUF utilizes the intrinsic manufacturing variations in a device to generate a fingerprint of the hardware that offers the valuable advantage of unclonability. This means that the device cannot be cloned even when a hacker has physical access to the device. Therefore, the PUFs are unique to their device and can be used as a security primitive to enable device-based identification, and authentication.
More importantly, PUFs can provide a low cost alternative solution for on-demand generation of  cryptographic keys from the device rather than the conventional methods, where the secret keys are produced and distributed by the server and stored in the IoT device memories {\cite{chatterjee2018building}}. 
 
%\textcolor{magenta}{this statement is not strong enough to describe the contributions of the work, I made an attempt to define the project but you need to improve it} 
In this paper, we developed a proof-of-concept key generation mechanism using static random-access memory (SRAM) PUFs, which does not involve the key storage in the devices' memory. This developed technology utilizes the idea of ternary PUFs \cite{cambou2018ternary} %\textcolor{magenta}{add references} 
that has proven to create stronger PUFs compared to common binary PUFs. The developed key generation mechanism has been tested using different SRAM devices. We also evaluated the ability of various fuzzy extractors proposed in the literature to extract the key of the noisy PUF input.

The rest of this paper is organized as follows: In section \ref{section:PUF}, we introduce the concepts of PUFs and focus on SRAM based PUFs and its characterization. In section \ref{sec:FuzzyExtractors}, we explain, how Fuzzy extractors can be used to correct the noisy input PUF response. In section \ref{sec:KeyGenFuzzyExtractors}, we describe the key generation and regeneration process using fuzzy extractors proposed in the literature. In section \ref{sec:Experiment}, we discuss our experimental setup, the custom-made software and Arduino shield which allows us to extract SRAM responses on demand. We then use these responses to generate secret keys using Fuzzy extractors.

\section{Physically Unclonable Functions}\label{section:PUF}
\subsection{PUF-based technology}
Physical unclonable functions (PUF) are the equivalent of human fingerprints; the variations created during fabrication makes each PUF authenticable from each other. Authentication protocols based on PUFs \cite{pappu2002physical,gassend2002silicon,gao2016emerging,herder2014physical,maes2010physically,jin2015introduction,delavor2014puf,guajardo2007physical,plusquellic2015systems}, can be effective when these PUFs show intra-PUF stability, and offer inter-PUF randomness. Memory structures \cite{holcomb2009power,maes2009soft,christensen2012implementing,prabhu2011extracting,chen2015comprehensive,cambou2017ag,zhu2015physically,vatajelu2016stt}, SRAM \cite{holcomb2009power,maes2009soft}, DRAM \cite{christensen2012implementing}, Flash \cite{prabhu2011extracting}, ReRAM \cite{chen2015comprehensive,cambou2017ag}, and MRAM \cite{zhu2015physically,vatajelu2016stt}, are all suitable to generate strong PUFs.

One method to generate PUFs from memory devices is to characterize a particular parameter $\mathcal{P}$ of the cells of its array. The values of parameter $\mathcal{P}$ vary cell to cell, and follow a distribution with a median value T. The cells with $\mathcal{P} < T$ generate ``0" states, and  others generate ``1" states. The PUF ``Challenges", also called ``initial" responses” is the data stream generated during ``enrollment" of the client’s devices by the server.  In some protocols, the Challenges can be generated by processing, or averaging, multiple queries and measurements of the PUFs. These Challenges can be represented by binary, ternary, or any other radix. 

During enrollment, the ``Challenges" of the PUFs located in each client device are securely downloaded in a data base of the server. The PUF “responses” is the data streams generated by the PUF during the life of the client devices. These PUFs are physical elements that can age, and be subject to temperature changes, electro-magnetic interferences, and other environmental effects. “Challenge-response-pairs” (CRP) are generated on demand by the server of  the PUFs of  client devices. When the PUF is a strong PUF,  the server needs to send “instructions” to the client devices to find the particular “address” in the PUF,  to generate challenge-response-pairs. The “Helper” is the data stream generated by the server from these challenges which is transmitted to the client devices to correct the responses. Some authors call the “instructions” the “challenges”, the CRPs then become the pairing of the initial responses to the freshly generated the responses.

When the CRP error rates are below 10\%, the responses can be used as part of the authentication protocol, which has significant commercial value to protect cyber physical systems. Satisfactory authentications have low false rejection rates (FRR), and low false acceptance rates (FAR). The use of PUFs to generate cryptographic keys from the responses is more challenging than generating responses for authentication; a single-bit mismatch in a cryptographic key is not acceptable for most encryption protocols. This requires schemes that fully correct the responses of the PUF.

\begin{figure}
    \centering
    \includegraphics[width = \columnwidth]{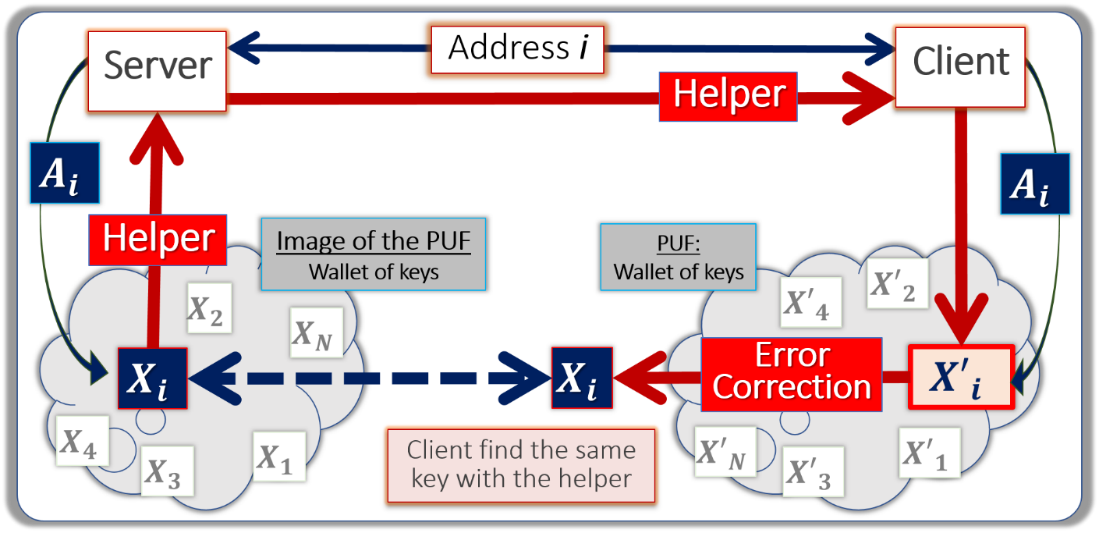}
    \caption{PUF-based cryptography with an error correcting scheme.}
    \label{fig:PUFbasedCryptography}
\end{figure}

A typical architecture to drive a network of client devices with PUFs is shown in Fig.\ref{fig:PUFbasedCryptography}. The server initiates the process by sending instructions to the client device to find the addresses \textbf{\textit{i}} within the PUFs which act as “wallet of keys”, and to extract the Responses \textit{\textbf{$X’_i$}}. The server independently analyses the challenges \textit{\textbf{$X_i$}} stored in the table at address \textbf{\textit{i}}, and generates the Helper with error correcting methods. The Helpers are transmitted to the client devices as part of the protocol. The client device corrects the response \textit{\textbf{$X’_i$}} with the Helper and correcting methods such as a fuzzy extractors. To be acceptable for encryption, the corrected responses, and the challenges of the server should be perfectly identical. Thereby, the same key \textit{\textbf{$X_i$}} is independently generated for encryption schemes.

Such PUF-based protocols have several weaknesses when utilized in IoT networks. The client devices are burdened, and need to consume additional computing power to run the error correcting codes; and such protocols increase the vulnerability to side channel attacks, differential power analysis, and the potential exposure of the Helpers. In this paper, we develop a PUF-based key generation scheme based on using a ternary state key generation method to exclude fuzzy cells in the PUF. We also utilize low power correction schemes to minimize the length of the Helper, and computing power needed to deliver error free keys to address the needs of a key generation mechanism in IoT networks.

\subsection{SRAM-based PUFs}
SRAM-based PUFs exploit the lack of perfect symmetry of their cells, which are designed with flip-flop logic. The SRAM cells tested in this work have six transistors per cell, four transistors for the flip-flop, one transistor to control the programming of the cell, one transistor to control the read. During the programming cycle, the flip-flop is set at ``0", or ``1"; during read cycle, the state of each flip-flop is detected. Power-off/power-on cycles are used to get PUF functionality from SRAM devices. During power-off/power-on cycles, the cells have an equal opportunity to recover a “0” or “1” state; however, the majority of the cells of the SRAM arrays have predictable behavior. In most cases, they have a preferred state either a “0” or a “1” due to natural variations, introduced during the manufacturing of the devices; the flip-flop cells are never perfectly symmetrical. Examples of variations include slight differences in the size of the transistors due to non-uniformities of the photolithography of the process, micro-defects, natural variations of the chemical composition of the metallic lines, and many other small fabrication parameters. 

Each SRAM cell is different from other cells; each SRAM device is in general, different from the others. When 256 cells are randomly selected, and subjected to power-off/power-on cycles, the vast majority of the data stream stored in these 256 cells is predictable, and can be used as a PUF challenge-response-pair. While the asymmetry is strong enough for most of the cells, the CRP error rates of SRAM PUFs can be prohibitive because some cells are too close to a symmetrical configuration, and they can flip randomly on both sides.

\subsection{Characterization of SRAM PUFs}
\begin{figure}
    \centering
    \includegraphics[width = \columnwidth]{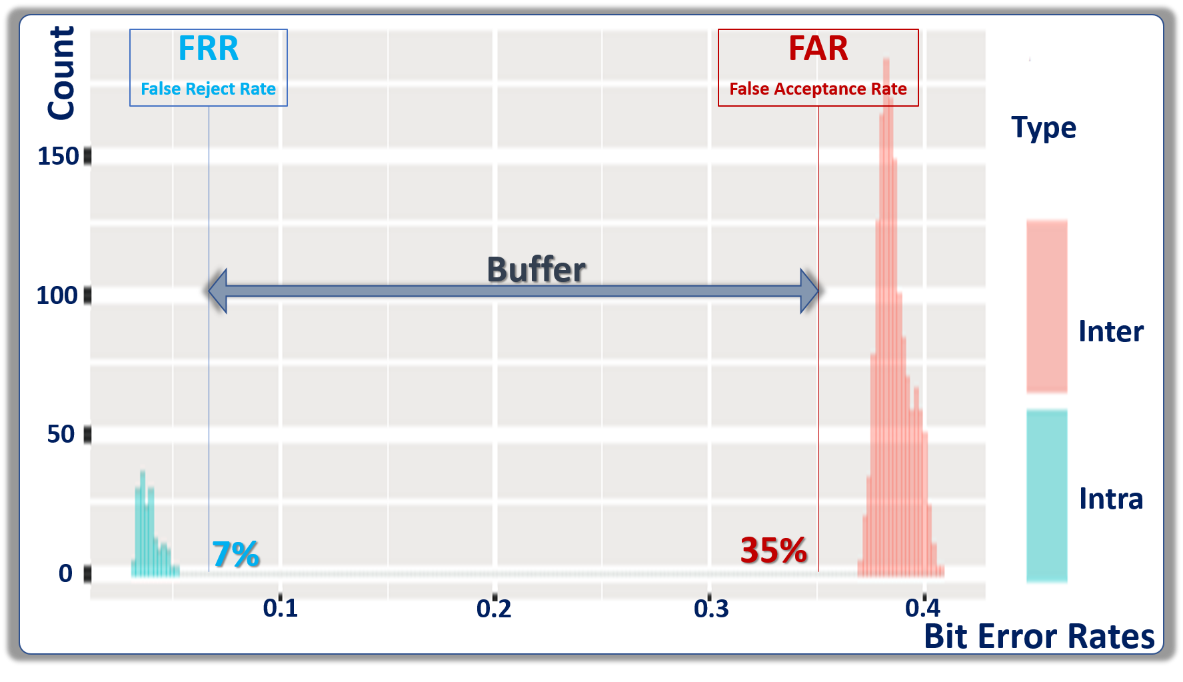}
    \caption{Characterization of the intra-PUF, and inter-PUF error rates.}
    \label{fig:characterizationPUF}
\end{figure}

The experimental work to analyze the CRP error rates is based on 45 different 32kByte commercial SRAMs, produced by Cypress Semiconductor. The SRAMs with the circuitry needed for power-off/power-on cycles are assembled in custom PCB boards, driven by Arduino boards. A complete power-off of the SRAMs is needed in spite of capacitive and inductive effects, which can prevent the power from switching off. To mitigate the problem, all SRAM I/Os are shorted to the ground during these cycles to drain the charges, and during periods of time that are long enough, one second in this case. The intra-PUF and inter-PUF error rated of these SRAM devices is summarized in Fig.\ref{fig:characterizationPUF}. The intra-PUF error rate is defined as the CRP error rates of each cell compared to itself. The inter-PUF error rate is defined as the CRP error rates of each cell compared to the cells located at the same address on a different SRAM. For this analysis, 10 devices were characterized. When used for authentication, this allows the modelling of FRRs, and FARs

\begin{figure}
     \centering
     \includegraphics[width = \columnwidth]{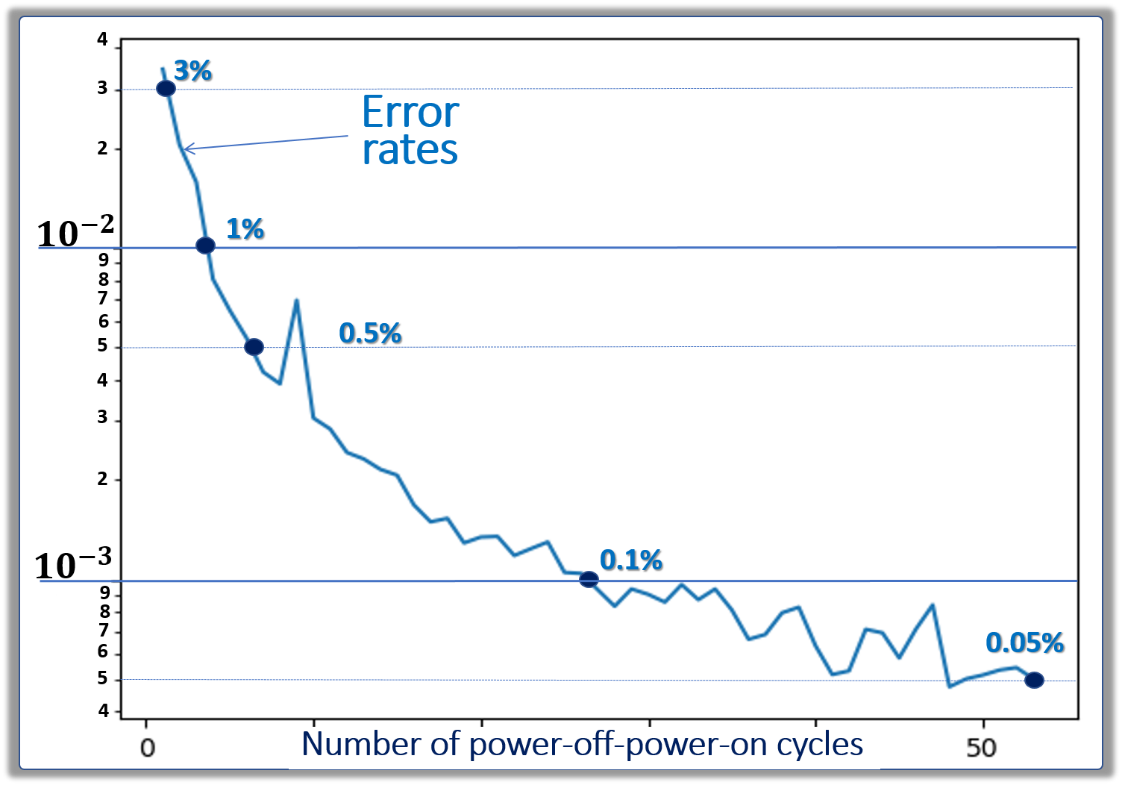}
     \caption{Error rate reduction by eliminating the fuzzy cells after power-off/wait state/power-on cycles.}
     \label{fig:errorRateReduction}
 \end{figure}

 As shown in Fig.\ref{fig:errorRateReduction}, the error rates of the SRAM PUFs are reduced with an enrollment that incorporates successive power-off/power-on cycles, to remove the erratic cells from the population. During each cycle, the cells that are flipping between “0”, and “1” are removed. The estimated error rates of the remaining cells and plotted in Fig. 3. Initially, before eliminating the weak cells, the CRP error rates are in the 3 to 5\% range. After four cycles the CRP error rates are reduced to about 1\%. After 25 cycles the CRP error rate drops below 0.1\%; then the reduction is slower. About 50 cycles are needed to get 0.05\% error rate, and 1000 cycles to get 0.01\%. Such error rates are still too high for cryptography. In the experimental work presented in this paper, the error rate before error correcting schemes can be adjusted from 3\% to 0.01\%, to study the respective efficiency of the schemes.
\section{Fuzzy Extractors}
\label{sec:FuzzyExtractors}
A fuzzy extractor will extract a uniformly random string $R$ and non-secret string $P$ (helper data) from its initial input $w$. This mechanism will allow the string $R$ which can be used as a key, to be reproduced exactly with the help of $P$, even though the input changes to some $w'$ but remains close to $w$. Fuzzy extractors are said to be \textit{information-theoretically secure} i.e., a crypto-system whose security is derived only from information theory, where a hacker will not have enough information to break the encryption, allows them to be used in cryptography.
\newline
%*********************************************Figure
\begin{figure}[!htbp]
\begin{center}
\centering
\includegraphics[width=\columnwidth]{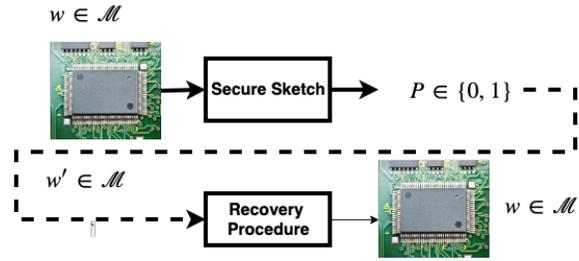}
    \caption{Secure Sketch}
    \label{fig: fuzzyExtractor_a}
    \end{center}
\end{figure}
%*********************************************Figure
Fuzzy extractors are constructed using \textit{Secure Sketch} (SS), which is a pair of randomized procedures ``sketch'' and ``recover'' which will allow precise reconstruction of the initial input from noisy input by making use of some helper data $P$.

In the ``sketch" phase, \textit{Helper Data} $P$ is extracted from initial input $w$, which can be made publicly available. This output $P$ will be used in the ``recover" phase along with noisy input $w'$ to recover $w$. This method is secure as the publicly available \textit{Helper Data} reveals little to no information about $w$. Fig. \ref{fig: fuzzyExtractor_a} summarizes the description of a secure sketch. 

%*********************************************Figure
\begin{figure}[!htbp]
\begin{center}
	\includegraphics[width=\columnwidth]{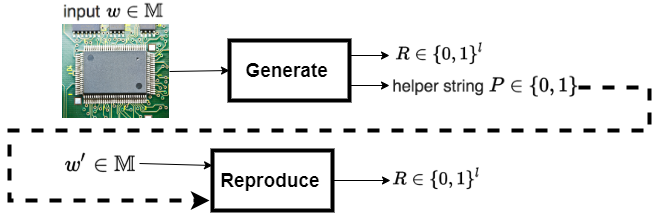}
    \caption{Fuzzy Extractor }
    \label{fig: fuzzyExtractor_b}
    \end{center}
\end{figure}
%*********************************************Figure

\begin{figure}[!htbp]
	\includegraphics[width=\columnwidth]{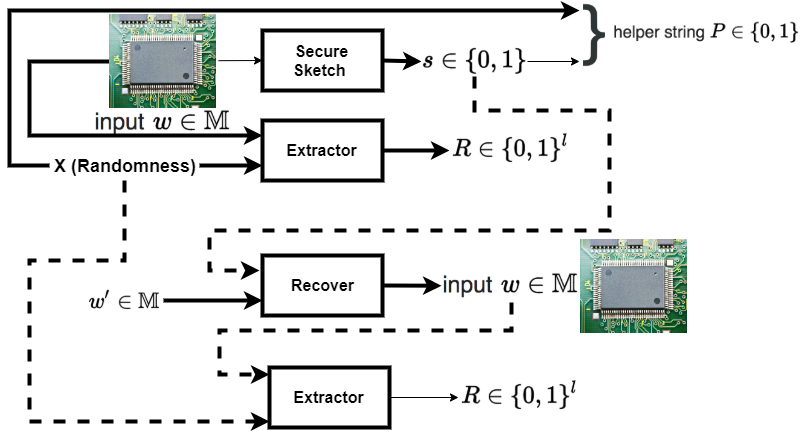} 
    \caption{ Construction of a fuzzy extractor using secure sketch extractor}
    \label{fuzzyExtractorConstruction}
\end{figure}

Fuzzy extractor is defined by a pair of randomized procedures ``generate'' and ``reproduce''. In the ``generate" phase, the fuzzy extractor uses the ``sketch'' phase of the SS where \textit{Helper data}, $P$ and \textit{Key}, $R$ from the given input $w$. The ``reproduce" phase uses the ``recover'' phase of the secure sketch which makes use of the \textit{Helper data} to recover the original input $w$ from a noisy input $w'$ along with the random extractor used in the ``sketch" phase, to extract the randomness from the recovered $w$. The ability to recover $w$ from $w'$ is highly dependent on the technique, usually an error correction scheme, used in the ``sketch'' phase of the Fuzzy Extractor. If the distance between the noisy input $w'$ and input $w$ is too large, it may not be possible to recover $w$ from $w'$. Fig. \ref{fuzzyExtractorConstruction} shows the construction of a Fuzzy Extractor using a secure sketch.  
%*********************************************Figure
%*********************************************Figure
Due to the error tolerance capability of Secure Sketches, their construction is based on error correcting codes. The error correcting code $C$ is used to correct errors in $w'$, even though  $w'$ may not be in $C$, by shifting the codeword. Two different constructions are used for secure sketch are provided \cite{dodis2004fuzzy} : 
\begin{itemize}
\item{Code-Offset Construction: For input $w$, select a uniformly random codeword $c \in C$, and set SS(w) to be the shift needed to get from $c$ to $w$: $SS(w)=w-c$. To compute $Rec(w',s)$, subtract the shift $s$ from $w'$ to get $c'= w'-s$: decode $c'$ to get $c$ and compute $w$ by shifting back to get $w=c+s$. When code $C$ is linear, the information in s is essentially the syndrome of w.}
\item{Syndrome Construction: The sketch SS(w) computes $s= \textit{syn}(w)$, where \textit{syn} is the syndrome. To recover the key, a unique vector $e$ is chosen such that $syn(e)= syn(w')-s$ and output $w=w'-e$.}
\end{itemize}

\section{Generation of cryptographic keys from PUFs using Fuzzy Extractors} 

\label{sec:KeyGenFuzzyExtractors}
Secret key generation using PUFs will allow users to produce a key from their own device which need not be stored in the devices' memory. Using PUFs to produce keys, will make the device unclonable and hence less susceptible to hacking. Using PUFs eliminates the complications and security issues related to key storage and distribution. Different PUFs have been used in the past to generate reliable and reproducible cryptographic keys using fuzzy extractors.
\newline
Different schemes have been proposed in the literature for generating reproducible keys using PUFs \cite{suh2007physical, kang2014performance,kursawe2009reconfigurable,ziola2014authentication,vskoric2005robust, lim2005extracting}.%, kang2014performance, kursawe2009reconfigurable, ziola2014authentication, vskoric2005robust, lim2005extracting,  kang2013implementation, delvaux2016efficient, taniguchi2013stable, price2014generate}. 
 Key generation scheme proposed in \cite{kang2014performance} uses BCH codes and random number generators for the construction of fuzzy extractors. The sole purpose of the BCH codes was to help reconstruct the PUF estimate from noisy PUF data, therefore, it will serve in the ``secure sketch" phase of the Fuzzy extractor. 
This scheme is one of the primitive schemes and does not require any complex decoders, hence the error correction capability is not good. This scheme may be good for authentication where a certain error margin is tolerated. In key generation schemes, where the key will be used for encrypting or decrypting the data, the error in reproducing the key has to be zero. 

%*********************************************Table
\begin{table*}[ht]
\caption{Comparison Of different fuzzy extractor schemes proposed in the literature.}
	\label{fuzzyTable}
		\begin{tabular}{|c|c|c|c|c|}
			\hline
			Fuzzy Extractor Construction   & Key Length    & Helper Data bits           & Failure Probability  & Flipping probability          \\ 
			\hline
			Reed Muller Generalized Multiple Concatenated coding	\cite{maes2009low}	  &128    & 13952   & $10^{-6}$ & 15\%  \\ 
			\hline
			BCH Repetition Code	\cite{maes2012pufky}	    &128  & 2052   & $10^{-9}$ & 13\%      \\ \hline
			Generalized Concatenated (GC) Reed Muller\cite{puchinger2015error} 		& 2048     & 2048   & $5.37. 10^{-10}$ & 14\%     \\
			 \hline
			GC Reed Solomon\cite{puchinger2015error}                         &1024 & 1024 &  $3.47. 10^{-10}$ & 14\% \\
			 \hline
			Polar Codes SC \cite{chen2017high}       &128                 & 896             &  $ 10^{-6}$     &15\%       \\ 
			\hline
			HA SCL Polar Codes \cite{chen2017high}       &128                  & 896             &  $ 10^{-9}$     &15\%       \\ 
			\hline
 			{Serially Concatenated BCH and Polar codes using SC decoder \cite{Korenda_IWCMC}}    & {250 }                         & {262}          & $ {10^{-8
 			}}$  & 15\%        \\ 
             \hline
			{Serially Concatenated BCH and Polar codes using Belief Propagation decoder \cite{Korenda_IWCMC}}     & {250 }                         & {262}          & $ {10^{-10
			}}$  & 15\%        \\ \hline
		\end{tabular}
        
\vspace{-0.5 cm}
\end{table*}
%*********************************************Table
Different fuzzy extractor architectures have been proposed in the literature, whose performances are compared in the Table \ref{fuzzyTable}.
In a recent work, a fuzzy extractor structure based on Polar codes for SRAM PUFs was proposed \cite{chen2017high}. This work utilized complex Hash-Aided syndrome construction (SC) decoder to ensure that the key was reproducible. The results showed that the key was reproducible with a failure probability of $10^{-9}$ and utilized 896 helper bits for a key of 128 bits. This work utilized polar codes for extraction of key and helper data, thus utilizing the property of zero mutual information between the helper and key. In our previous work, we utilized serially concatenated polar and BCH codes to generate 250 bit keys with 262 helper data bits using SC decoder and Belief propagation decoders, while maintaining a comparable failure probability of $10^{-8}$ and $10^{-10}$, respectively \cite{Korenda_IWCMC}.   %\textcolor{magenta}{add some sort of evaluation statement to justify why you mentioned their work, only because it's SRAM? it has the best result so far? it's good but we have ternary PUF that improve the performance} \textcolor{blue}{what is your method that you refer to it in the table, it needs to be discussed and described here, see my comment in magenta in the experimental results section about the paragraph you need to add here}

Here, we provide a proof of concept of key generation using SRAM using ternary state and compare the performance of the current state-of-the-art generators to see what is the trade-off between good error correction to achieve zero error key regeneration with the chance of getting false positive for a PUF which is not ideal in terms of intra-PUF distance. We would also like to determine the trade-off between the performance, the level of computations and delay, and the number of helper data bits needed to regenerate the key from Noisy PUF responses. The device can be utilized to improve the raw probability of error for each cell using the ternary state method. We can realise the need to use sophisticated error correction coding that will sufficiently correct the input bits, without increasing the false authentication rate of the device.  

\section{Experimental Results} \label{sec:Experiment}
	As a proof of concept, the Cypress CY62256N was chosen as the SRAM to read from. It can store up to 32 KB of data, which was deemed more than sufficient source of entropy. A custom Arduino Mega 2560 shield was designed for the DIP variant to allow for easily switching between different packages. Every pin of the package is also connected to a toggleable switch such that a logical OFF corresponds to ground; without this design, the SRAM cannot be automatically controlled to dissipate the current.
	
	First, the Arduino is programmed with three modes in mind: the first mode, upon receiving a special flag over the serial port over USB, reads the entire SRAM one byte at time, sends each back over the serial connection, and powers it down; the second mode, when receiving a different flag, will manually power the SRAM down; the third and last mode, when receiving neither of the former two flags, will interpret the value as a bit address and send back the read bit as the smallest possible word (in this case a byte). The last mode assumes to continuously read a series of bit addresses, until the second mode is triggered and powers off the SRAM to correspond to the PUF’s lifetime.
	
	A Python script communicates with the Arduino over the serial connection. By default, it indefinitely loops, requests the Arduino for the next PUF, and saves the entire contents of the SRAM onto a successive binary file. A separate script is then used to read every binary file and iterate over every cell based on the following criteria: if a cell's value has changed in any of the reads, it is from then on marked as an ``X" to represent its unstable nature. If that cell has remained the same value (either 0 or 1) after iterating through every read, then that cell is given that value in the final enrollment. One can also tell the script to be less stringent, and instead set the final values based on the number of times the cell's value was 1 over the total number of reads. For instance, if the cell was 1 less than 30\% of the reads, it is marked with ``0"; if it was 1 more than 70\% of the reads, then it is marked with a ``1"; otherwise, it is marked with an ``X". For the purpose of this experiment, the cell prone to remain constant at least 46\%, it is marked with the constant ; otherwise it is marked with a ``X".
	%\colorbox{yellow}{which one you did here, what is your threshold? }

\begin{figure}[!htbp]
\begin{center}
\centering
\includegraphics[width=\columnwidth]{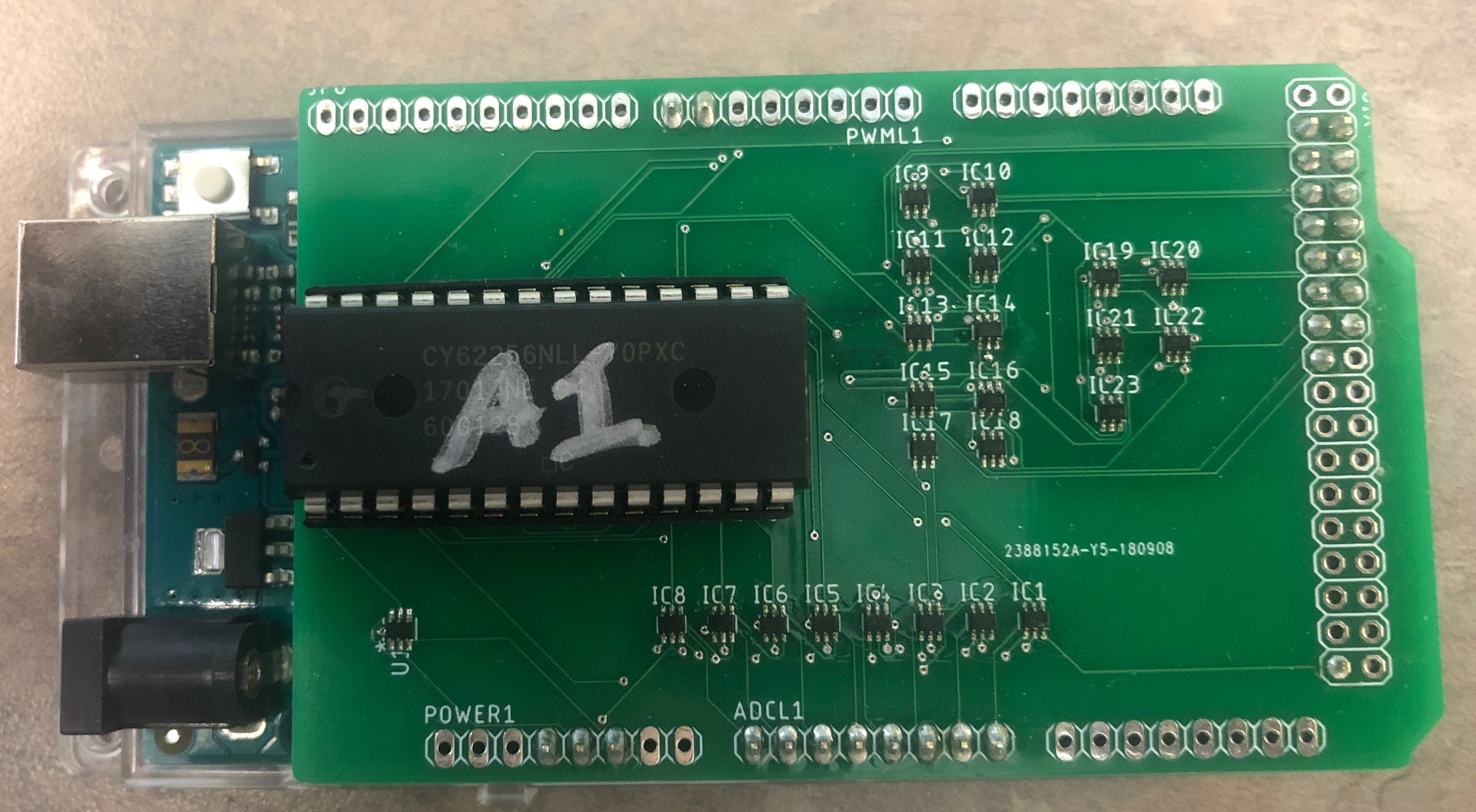}
    \caption{SRAM Device}
    \label{fig: SRAMdevice}
    \end{center}
\end{figure}

%\textcolor{magenta}{Move this to the last paragraph of the previous section and add a stronger statement to describe the goal of the work, you want to offer a proof of concept of key generation using SRAM using ternary state and compare the performance of the current state-of-the-art generators to see what is the trade-off between good error correction to achieve zero error key regeneration with the chance of getting false positive for a PUF which is not idead in terms of intra-puf. Another goal would be the tradeoff between the performance, the level of computations and delay, and the number of helper data you need to made available. One of your claims could be with improving the raw probablibilty of error for each cell using the ternary state method, you can realse the need to use sophisticated error correction coding that .... } 
We tested the various fuzzy extractor schemes in the literature using an SRAM PUF device produced by the NAU Cybersecurity lab shown in Fig. \ref{fig: SRAMdevice}.

%*********************************************Figure
%*********************************************Figure
 The SRAM device has a software package which allows us to interact with the device. It has various functions embedded in it, some of which include, reading the real time PUF response after refreshing the SRAM every 2 seconds, counting the number of mismatches between every response, enrolling the PUF, where it utilizes 100 reads of the PUF to determine the stable cells in the SRAM device and stores the challenge in the server.  
\newline
We use a technique called addressable PUF generator (APG), which allows us to extract a unique fingerprint from the device, by randomly selecting the cells which will be used as the fingerprint input to the fuzzy extractor. 

\subsection{Addressable PUF Generator}
\begin{figure}[bh!]
    \centering
    \includegraphics[width=\columnwidth]{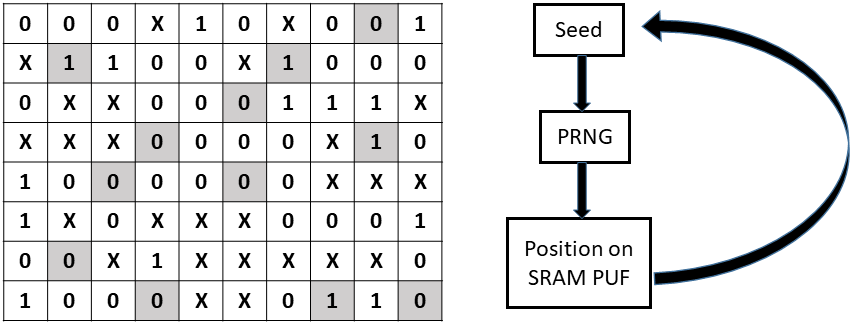}
    \caption{Using Pseudo random number generator (PRNG) to select bits from the finger print of the SRAM Device}
    \label{fig:position}
\end{figure}

We use the technique of APG described in \cite{cambou2018encoding} to generate challenges for our PUF devices. In APG technique, a random  number and the user ID are XORed and given as an input to a hash function. The hash digest is used to identify a location in the PUF. 
This location is given as an input to a pseudo random number generator, which will allow us to pick different memory cells from the device, giving a range of cells, which are randomly selected from the entire memory chip. Fig. \ref{fig:position}, shows various cells selected in an SRAM fingerprint, using a random seed given to a pseudo random number generator (PRNG), allowing us to use bits spread across the finger print to extract the challenge and response.

\subsection{Varying the error rate of the device}
We use ternary PUF concept to mask the bits which are easily prone to flip in the memory device. The software allows us to manage the error rate by varying the number of reads used to average the SRAM challenge. We can mark the bits which are easily prone to flip with an ``X". More number of reads used will lead to a more stable challenge as the bits prone to flip over a large number of reads will be marked with an ``X". By this method, we can determine the cells which are able to produce stable ``0" and ``1" and eliminate cells with the ``X".  For example, the error rate when only 10 reads are used to average the challenge is higher when compared to using 100 reads to average the challenge.
\begin{table*}[h]
    \centering
    \caption{Effect of Intra and Inter PUF distance on Key generation using different Fuzzy extractors}
    \begin{tabular}{|c|c|c|}
    \hline
       \textbf{ Fuzzy Extractor} &\textbf{ \% Error when the same PUF is used } &\textbf{ \% Error when different PUF is used}\\
        \hline
       BCH Fuzzy Extractor \cite{dodis2004fuzzy}& 0.2 \% & 47.62\% \\
       \hline
       Efficient Fuzzy Extractor using BCH \cite{kang2014cryptographie}& 12.5 \% & 48.98\%\\
       \hline
    Polar using HA-SCL \cite{chen2017high} & 0 \% & 19.23\%\\
    
    \hline 
    Polar using serially concatenated BCH and Polar Codes \cite{Korenda_IWCMC} & 0 \% & 23.68\% \\
    \hline
    
    \end{tabular}
    
    \label{tab:errorResult}
\end{table*}
\subsection{Generating keys and storing them in the server}
 We simulated the fuzzy extractors proposed in the literature \cite{maes2009low, maes2012pufky, puchinger2015error, chen2017high} in MATLAB. We utilize the fuzzy extractors to generate keys from challenges by varying their error rate by changing the number of reads used to average the challenges and mask the flipping bits with``X". We generate secret keys and helper data using the registration phase of the fuzzy extractor.
 \subsection{Regeneration of Keys using PUF response}
 Using the helper data produced in the key generation process, the key is regenerated. The ability of the Fuzzy extractor to regenerate the key under different error rates is evaluated.
 The False Authentication Rate and False Rejection Rate while using different Fuzzy extractors under various error rates is determined. 
\newline
We use real-time SRAM responses of the device along with the helper data produced in the key generation phase to regenerate the key. The average error when an SRAM PUF is used to generate keys using the fuzzy extractors described in  \cite{dodis2004fuzzy,kang2014cryptographie,chen2017high,Korenda_IWCMC} is reported in the Table \ref{tab:errorResult}. The FAR when a different user imposters as an authenticated user is also reported in the Table \ref{tab:errorResult}. To calculate the FAR, another PUFs response was utilized with the helper data provided by the registered challenge. The FAR was low when simpler fuzzy extractor methods like \cite{dodis2004fuzzy,kang2014cryptographie} were utilized, but the FRR was high as they were not strong enough to correct the errors in the Noisy PUF. When stronger methods like \cite{chen2017high,Korenda_IWCMC}  were utilized, the FAR was higher compared to the weaker methods, as they were capable enough to correct more errors in the PUF, thereby reducing their FRR.  

\begin{figure}[bh!]
    \centering
    \includegraphics[width=\columnwidth]{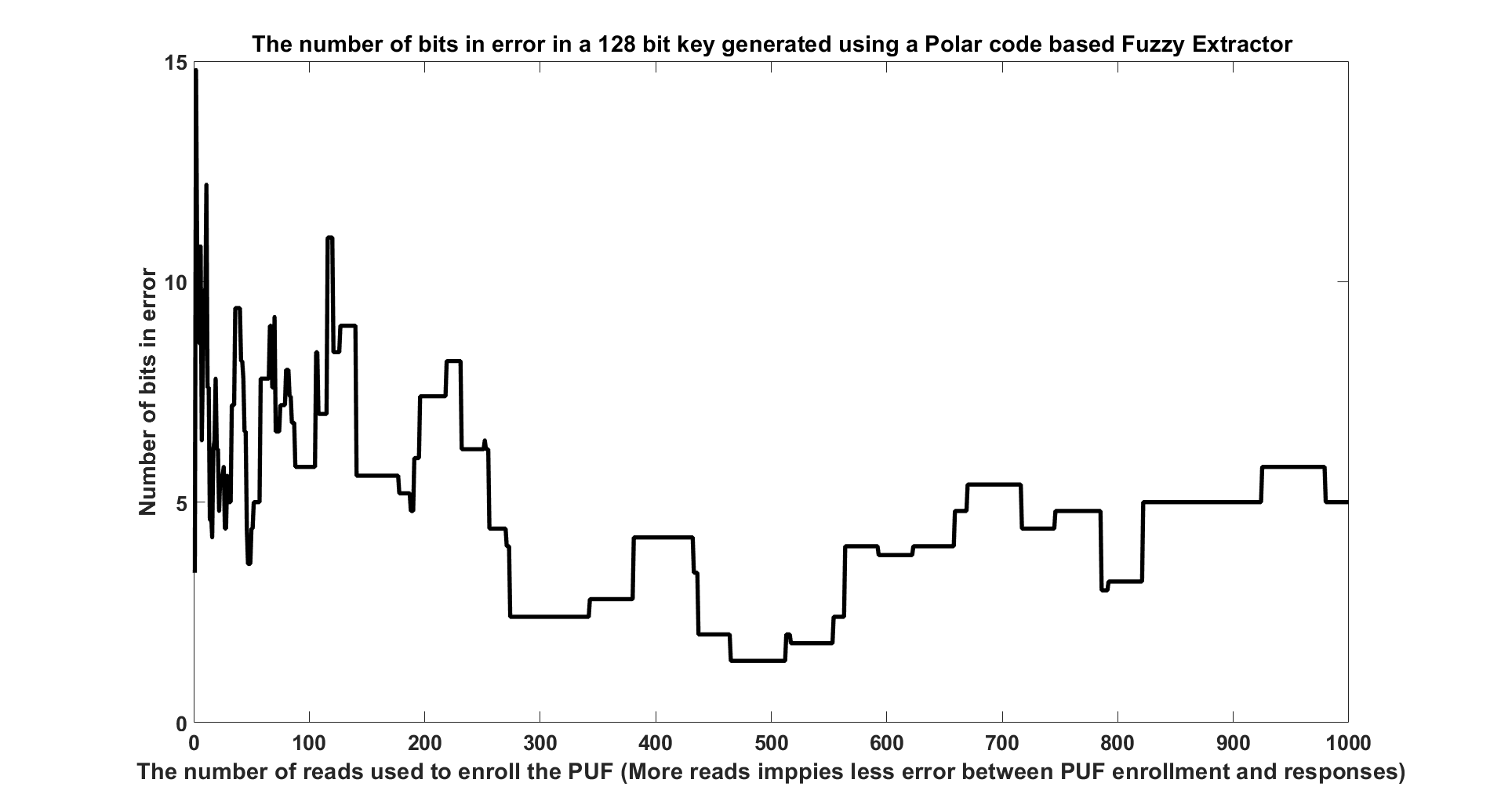}
    \caption{\% Error in generating 128 bit keys from a Polar code based Fuzzy Extractor.}
    \label{fig:averageBER}
\end{figure}

In Fig. \ref{fig:averageBER}, shows the ability of a Polar based Fuzzy extractor to derive the key at different error rates. The error rate of the SRAM is varied by using the ternary technique. Here, we used different number of cycles/reads to calculate the enrolled challenge, stored in the server. More number of bits will be marked with an ``X" when more number of PUF responses are used to enroll the challenge. Marking the cell with an ``X" indicates, removing the cell which flipped during one of the responses used to make the enrolled challenge to be stored in the server, thereby reducing the error of the PUF. The error in retrieving the key from Noisy SRAM responses using a fuzzy extractor was calculated for different error rates of the PUF.  We observed the same pattern noted in Fig. \ref{fig:averageBER} for 15 different SRAM devices. There is a clear decline in the error rate of the key as the error in the PUF reduces. The error in key tends to increase even when the PUF error is very low. This may be due to the fact that the error correction mechanism used in the Fuzzy extractor is over correcting the data, thereby changing the correct bits. This may be one of the factor for increase in the error of the key, even after decrease in the PUF error. Other factors may include different parameters in the error correction code. This is an ongoing work, therefore we will be working on testing each of the parameters of the error correction codes to check its effect on the key generation scheme.

%\textcolor{magenta}{let me know when you are ready for me to read this part, as I said the main reasons for having this paper is still not mentioned anywhere }

\section{Conclusions}
In this paper we developed a proof of concept for a practical key generation scheme in IoT networks that does not rely on storing the private keys in the memory of the IoT devices, but rather generates a private key from the unique fingerprints of the embedded memory of the IoT devices on an on-demand basis. Such method can add another layer of security to the common cryptographic techniques (e.g., PKI) from physical attacks. Therefore, this method can offer a scalable security solution for large-scale IoT networks without the need to generate, distribute and store the private keys in billions of IoT devices. 

We have utilized a recent idea of ternary-state PUFs rather than the traditional binary PUFs to identify the more reliable cells in the memory of the device that can substantially increase the robustness of the PUF responses for a very high sensitive application such as generating the private keys for encryption. The performance of this method has been  tested for several SRAM chips using our device. We can vary the average error rate of PUF (i.e., selecting more stable cells which have lower probability of flipping) using different thresholds to define the ternary states, thus providing more stable PUF responses as needed. These responses can further be used to produce secret keys using fuzzy extractors. We tested the efficiency of the fuzzy extractors using real time SRAM PUF responses while varying its error to test the false authentication and false rejection rate. In summary, we like to note that the since SRAM PUFs depend on the flip flop logic and lack perfect symmetry, it is considered as a weaker PUF compared to some other memory-based PUFs (e.g. ReRAM based PUF), however, it's a common NVM technology available in several IoT devices and offers an easy and accessible technology to explore. 
\bibliographystyle{IEEEtran}
\bibliography{IEEEabrv,main}
\end{document}